\documentclass[showpacs,preprintnumbers,twocolumn,superscriptaddress]{revtex4-1}
\usepackage[dvipdfmx]{graphicx}
\usepackage{amsmath,amssymb}
\usepackage{bm}

\begin{document}

\title{
Possible generation of anomalously soft quark excitations
at nonzero temperature:\,
Nonhyperbolic dispersion of parapion and van Hove singularity
}

\author{Masakiyo Kitazawa}
\affiliation{Department of Physics, Osaka University, Toyonaka, 
Osaka 560-0043, Japan}
\author{Teiji Kunihiro}
\affiliation{Department of Physics, Kyoto University, Kyoto 606-8502, Japan}
\author{Yukio Nemoto}
\affiliation{Department of Physiology, 
St. Marianna University School of Medicine, Kawasaki, Kanagawa 216-8511, Japan}

\begin{abstract}

We study the quark spectrum at finite temperature near and 
above the pseudocritical temperature of the chiral phase 
transition incorporating the effects of the collective  modes 
with the quantum number of the sigma (parasigma) and pion (parapion)
in a chiral effective model with a nonzero current quark mass.
Below the pion zero-binding temperature where the pionic modes 
are bound, the quark self-energy has van Hove singularity 
induced by the scattering of quarks with 
the composite bound pions with a nonhyperbolic dispersion curve.
This singularity is found to cause a drastic change in the quark spectrum 
from that in the mean field picture near the pseudocritical temperature:
The quark spectrum has an unexpected sharp peak at an energy 
considerably lower than the constituent quark mass,
while the spectrum approaches the mean field one at high temperatures.
We clarify that  the emergence of this anomalous structure of the quark spectral function 
originates from the composite nature of the pionic modes 
with a non-Lorentz invariant dispersion relation in the medium at finite temperature.
\end{abstract} 
\date{\today}
\maketitle

\section{Introduction}
\label{sec:intro}

The exploration of the nature of the hot medium 
near the phase boundary of chiral and deconfinement phase 
transitions is an intriguing subject in quantum chromodynamics (QCD).
Experimental results in heavy-ion collisions at the 
Relativistic Heavy Ion Collider (RHIC) \cite{RHIC} and 
the Large Hadron Collider (LHC) \cite{LHC} suggest that 
the quark-gluon medium near the phase boundary is a strongly 
interacting system.
The properties of the hot medium are also actively 
investigated by the lattice QCD Monte Carlo simulations, 
which have recently revealed that the phase transition 
between hadronic and quark-gluon media at vanishing 
baryon chemical potential is a smooth crossover without 
a sharp boundary \cite{lat-Tc}.
The lattice simulations also suggest that the thermodynamic 
observables including higher-order fluctuations of conserved 
charges are well described by the hadron resonance gas 
model below the pseudocritical temperature $T_{\rm PC}$,
but such a picture breaks down in a narrow range of 
temperature ($T$) near $T_{\rm PC}$ \cite{lat-Tc,lat-fluc}.
This result indicates that, despite the crossover nature, 
the hot medium suddenly changes its character  
from that of a simple system composed of approximately free hadrons to a highly 
correlated system with unknown but intriguing degrees of freedom
in the vicinity of $T_{\rm PC}$.

To explore the nature of the hot medium 
above $T_{\rm PC}$, it is natural
to begin with an investigation of 
the existence and properties of collective excitations 
having the quantum numbers of the quarks and gluons.
As for collective modes carrying quark quantum number, 
it is notable that recent nonperturbative analyses on 
the quark spectral function 
on the lattice \cite{KK07,KK09,Kaczmarek:2012mb}, 
and Schwinger-Dyson approaches \cite{Harada:2008vk,
Muller:2010am,Mueller:2010ah,Qin:2010pc,Nakkagawa:2011ci} 
indicate the existence of such quasiparticle excitations 
even for temperatures not much greater than $T_{\rm PC}$.

For temperatures near but above $T_{\rm PC}$, interesting 
ingredients come into play owing to the strong coupling.
One of them is a possible existence of hadronic excitations 
that may survive the phase transition.
Indeed, lattice simulations show that charm quarkonia
can still exist as relatively stable states with an 
increasing width even well above $T_{\rm PC}$ \cite{charm}.
Another example of such hadronic states is the 
soft modes of chiral phase transition \cite{Hatsuda:1985eb}.
When the chiral transition is not so strong first order,
some specific collective modes of quarks and antiquarks 
have a chance to develop in the scalar ($\sigma$) and 
pseudoscalar ($\pi$) channels near the critical temperature 
in accordance with the 
enhancement of the fluctuations of the order parameter.
Moreover the masses (peak position of the spectral function) 
of these collective modes decrease as the system approaches the critical point, and 
 these modes are called the soft modes of chiral transition 
\cite{Hatsuda:1985eb}:
They become exactly massless at the critical 
temperature in the chiral limit.

When such soft modes exist above $T_{\rm PC}$, they 
can in turn affect the properties of the quasiquark excitations.
This possibility was explored in Ref.~\cite{KKN06} 
in a two-flavor Nambu--Jona-Lasinio(NJL)
model as in Ref.~\cite{Hatsuda:1985eb} in the chiral limit.
In this case, the chiral transition at nonzero $T$ is of 
second order and the quark has no constituent quark mass
above the critical temperature $T_c$, where
the well-developed soft modes appear near $T_c$.
It was shown that the fermion spectrum at low momentum
has a three-peak structure for $T\sim T_c$;
the fermion spectrum acquires a sharp peak at low energy in addition 
to normal and plasmino modes having thermal masses.

It is worth mentioning that the emergence of the three-peak structure in the fermion
 spectrum is a universal phenomenon for fermion-boson systems at nonzero temperature $T$
when the fermion mass $m_f$ is not so large \cite{KKN07,KKMN}; see also \cite{BBS}.
In Ref.~\cite{KKN07}, the fermion spectrum at nonzero $T$ was investigated 
in a simple Yukawa model composed of a massless fermion 
and an {\em elementary} boson with mass $m_b$
at the one-loop order, where the boson dispersion relation, $\omega_b=\omega_b(q)$,
is simply assumed to be of the hyperbolic form $\omega_b(q)=\sqrt{m_b^2+q^2}$
and the possible modification of it owing to the coupling to the fermion at $T\not=0$ is neglected.
It was found that the fermion spectrum at low momentum
has a three-peak structure for $T\simeq m_b$.
The existence of the sharp peak in the quark spectrum at low energy is later
confirmed in various models and analyses incorporating higher-order
contributions \cite{Harada:2007gg,Harada:2008vk,Qin:2010pc}, 
in some of which
the needed nonzero boson mass $m_b$  is supplied by the thermal mass.
On the other hand, it was shown in a Yukawa model with 
a massive fermion and an elementary massive boson
\cite{KKMN} that the nonzero fermion mass $m_f$
tends to suppress the appearance of the sharp peak at small 
energy that would be seen for $T\simeq m_b$: 
Such a peak can exist in the fermion spectrum only 
when the masses satisfy the condition $m_f \lesssim 0.2 m_b$.

The purpose of the present study is to extend the 
analysis in Ref.~\cite{KKN06} to the case off the 
chiral limit with nonzero current quark mass $m_0$.
With the explicit chiral symmetry breaking, 
the constituent quark mass takes nonzero 
values for all $T$, while the soft modes in the $\sigma$ 
and $\pi$ channels do not become massless. 
In view of the analysis in Ref.~\cite{KKMN} on the effect
of nonzero fermion mass and 
the fact that the nonzero mass of the bosonic modes
would also act to suppress the thermal effect 
on the fermion spectrum, one might suspect that the interesting 
structure in the quark spectrum obtained in the chiral 
limit will be blurred by nonzero $m_0$.
In this paper, we shall show that it is not the case.
One of the basic facts is the existence of the composite 
pionic mode with a  {\em stability} above $T_{\rm PC}$:
Because the constituent quark mass takes a nonzero value 
above $T_{\rm PC}$, the soft pionic modes can be stable 
against the decay into a quark and an antiquark even above 
$T_{\rm PC}$ up to some temperature at least one-loop level
\cite{Hatsuda:1994pi}.
We call the soft mode in the pionic channel existing above 
$T_{\rm PC}$  the parapion.
The other important ingredient leading to the results 
contrary to the naive suspect is a well-known fact 
that the dispersion relation $\omega_{\pi}(q)$ of the 
pionic mode in the medium at $T\not=0$ is generically  
different from the hyperbolic form given 
by $\omega_{\rm rel}(q)=\sqrt{q^2+[\omega_{\pi}(0)]^2}$
because of the violation of Lorentz symmetry at nonzero 
temperature and/or density \cite{Shuryak:1990,Pisarski:1996}.
We shall show that van Hove singularity 
\cite{vanHove,Brown1989,Shuryak:1990,CM} 
is brought about in the quark self-energy through the 
scattering of quarks with the pionic modes having such a 
modified dispersion relation, and the singularity 
drastically changes the quark spectrum.
In particular, we find that 
the quark spectral function has a sharp peak 
at an energy significantly lower than the constituent quark 
mass, which is quite reminiscent of but has a different origin 
from that of the peak found in Refs.~\cite{KKN06,KKN07,
Harada:2008vk,Harada:2007gg,Qin:2010pc}.
It will also be addressed that the modification of 
the quark spectrum with this mechanism is expected to take
place when a bosonic mode that couples to the quark 
has a nonhyperbolic dispersion relation irrespective of
the detailed structure of the dispersion relation.
Indeed, possible phenomenological 
consequences of such a modified dispersion relation of 
the pionic mode in the hot and/or dense medium were 
discussed in various contexts by many authors 
\cite{Brown1989,Shuryak:1990,Pisarski:1996}.

The paper is organized as follows.
The next section deals with the chiral soft modes.
In Sec.~\ref{sec:quark}, we calculate the quark self-energy 
due to the soft modes, and evaluate the quark spectral function.
The numerical results are shown in Sec.~\ref{sec:num}.
The final section is devoted to a summary and concluding remarks.

\section{Fluctuation modes}
\label{sec:fluctuation}

To study the fluctuation modes
in the scalar ($\sigma$) and pseudoscalar ($\pi$) 
channels on the spectral 
properties of quarks near the phase boundary, 
we employ the two-flavor NJL model \cite{Nambu:1961tp} as an effective model of 
low-energy QCD \cite{Hatsuda:1994pi}
\begin{align}
  \mathcal{L}=\bar{\psi} (i \partial \hspace{-0.5em} /  - m_0) \psi
  + G_S [(\bar{\psi} \psi)^2 + (\bar{\psi}i\gamma_5\bm{\tau}\psi)^2],
  \label{eq:NJL}
\end{align}
with $\bm{\tau}$ being the flavor SU(2) Pauli matrices
and the nonzero current quark mass $m_0=5.5$ MeV.
The coupling constant $G_S=5.5$ GeV${}^{-2}$ 
and the three-dimensional cutoff $\Lambda=631$ MeV are  
determined so as to reproduce the pion mass, the pion decay constant and the
quark condensate in vacuum \cite{Hatsuda:1994pi}.
We can expect that 
there will be no essential systematic uncertainty in the numerical results to be presented 
in the present work, once the parameters are fitted to
reproduce the physical values in vacuum, although another parameter set may be possible for that.

\begin{figure}[t]
\includegraphics[width=0.49\textwidth]{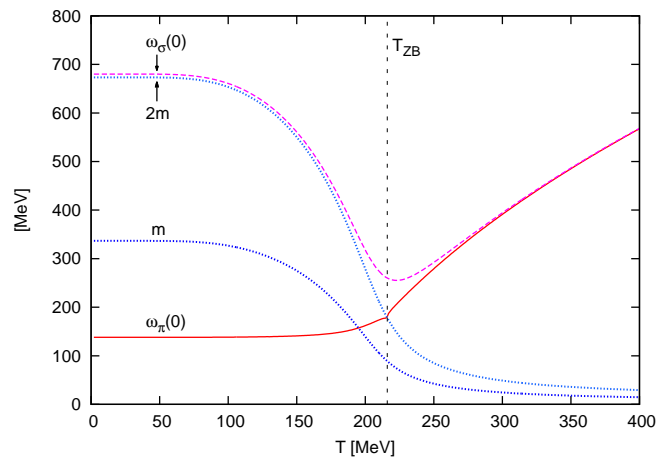}
\caption{Temperature dependence of the constituent quark mass $m$,
and the solution of Eq.~(\ref{eq:D=0}) at zero momentum for $\pi$ and $\sigma$ 
channels, $\omega_\pi(0)$ and $\omega_\sigma(0)$, respectively.
Twice the constituent quark mass is also plotted.
The vertical line shows the pion zero-binding temperature $T_{\rm ZB}$.
}
\label{fig:mass}
\end{figure}

The constituent quark mass in the self-consistent mean-field 
approximation (MFA) at $T\not=0$ is given by 
\begin{align}
m = m_0 - 2G_S \langle\bar\psi\psi\rangle,
\label{eq:m}
\end{align}
with the chiral condensate $\langle\bar\psi\psi\rangle$
evaluated with the mass $m$.
In Fig.~\ref{fig:mass}, we show the $T$ dependence of the resultant $m$.
The figure shows that in the vacuum $m=337$ MeV is 
significantly larger than $m_0$ as a consequence of the 
spontaneous chiral symmetry breaking.
For nonzero $T$, the constituent quark mass smoothly 
decreases in accordance with the chiral restoration in medium.
Because of the crossover nature, there are several 
definitions of the pseudocritical temperature $T_{\rm PC}$.
One can, for example, define $T_{\rm PC}$ as the 
temperature at which the magnitude of the chiral 
condensate becomes half the vacuum value. 
With this definition we have $T_{\rm PC}\simeq 192$ MeV.
Another possible choice is to use the dynamic chiral susceptibility
in the spacelike region \cite{KKNprep}, which diverges 
at the critical point when the transition is second order \cite{Fujii:2003bz}:
When $T_{\rm PC}$ is defined as the temperature where the 
dynamic chiral susceptibility with the momentum $|\bm{q}|=10$ MeV
has the maximum,
we have $T_{\rm PC}\simeq 206$ MeV.
One can also define $T_{\rm PC}$ as the
temperature at which the static chiral susceptibility has the maximum,
which gives $T_{\rm PC}\simeq 211$ MeV in this model.

\begin{figure}[t]
\includegraphics[width=0.49\textwidth]{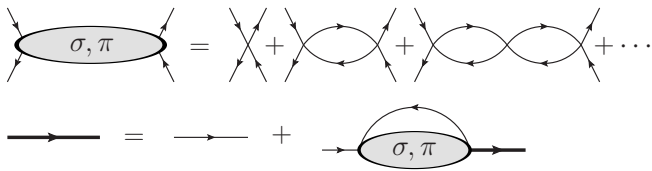}
\caption{
Diagrammatic representation of the  propagators of 
the bosonic and fermionic modes in this study.
The shaded area of the upper diagram
represents the propagator of  the sigma or pionic modes, 
$D^R_{\sigma (\pi)}(\bm{q},\omega_n)$ in Eq.~(\ref{eq:d-sp}), 
while the lower line represents the quark propagator 
defined in Eq.~(\ref{eq:sgtld}).
}
\label{fig:diagram}
\end{figure}

The properties of the fluctuation modes in the $\sigma$ and 
$\pi$ channels are encoded in the retarded propagator of 
these channels, $D^R_{\sigma}(\bm{q},q_0)$ and 
$D^R_{\pi}(\bm{q},q_0)$, respectively.
In the random phase approximation, 
these propagators are given by 
\begin{align}
  D^R_{\sigma(\pi)}(\bm{q},q_0) 
  = -\frac{2G_S}{1 + 2G_S Q^R_{\sigma(\pi)}(\bm{q},q_0)} ,
  \label{eq:d-sp}
\end{align}
with the one-loop quark-antiquark polarization functions
$Q^R_{\sigma(\pi)}(\bm{q},q_0)$.
The diagrammatic representation of Eq.~(\ref{eq:d-sp}) is 
shown in the upper part of Fig.~\ref{fig:diagram}.
The imaginary-time (Matsubara) propagators corresponding to 
$Q^R_{\sigma(\pi)}(\bm{q},q_0)$ are
\begin{align}
  {\cal Q}_{\sigma}(\bm{q},\nu_n) =& T\sum_m \int \frac{d^3p}{(2\pi)^3}
  {\rm Tr}[{\cal G}_0(\bm{p},\omega_m) \notag \\
  &\ \times {\cal G}_0
  (\bm{q}+\bm{p},\nu_n+\omega_m)], 
  \label{eq:Q-sig}  \\
  {\cal Q}_{\pi}(\bm{q},\nu_n) =& \frac{T}{3}\sum_m \int \frac{d^3p}{(2\pi)^3}
  {\rm Tr}[i\gamma_5 \bm{\tau} 
  {\cal G}_0(\bm{p},\omega_m)
\nonumber \\
  &\ \times i\gamma_5\bm{\tau}{\cal G}_0(\bm{q}+\bm{p},\nu_n+\omega_m)],
  \label{eq:Q-pi}
\end{align}
where $\mathcal{G}_0(\bm{p},\omega_n)=[i\omega_n \gamma_0
-\bm{p}\cdot\bm{\gamma}-m]^{-1}$ is the 
quark propagator in  the MFA, and $\nu_n=2n\pi T$ and 
$\omega_n=(2n+1)\pi T$ denote the Matsubara frequencies 
for bosons and fermions, respectively, and 
Tr denotes the trace over the color, flavor and Dirac indices.

After the summation of the Matsubara frequency 
analytically and the analytic 
continuation with a replacement $i\nu_n\to q_0+i\eta$ in 
Eqs.~(\ref{eq:Q-sig}) and (\ref{eq:Q-pi}), 
we obtain the corresponding retarded polarization functions 
$Q_\sigma^R ( \bm{q},q_0 )$ and $Q_\pi^R ( \bm{q},q_0 )$.
For the numerical calculation of $Q^R_{\sigma(\pi)}( \bm{q},q_0 )$, 
we first calculate their imaginary parts and then evaluate
the real parts with the Kramers-Kronig relation
\begin{equation}
  {\rm Re}Q^R_{\sigma(\pi)}(\bm{q},q_0) = 
  -\frac{1}{\pi}{\rm P}\int_{-\Lambda'}^{\Lambda'} dq_0' 
  \frac{{\rm Im}Q^R_{\sigma(\pi)}(\bm{q},q_0')}{q_0-q_0'} ,
\label{eq:KramersKronig}
\end{equation}
where P denotes the principal value. 
The cutoff of the $q_0'$ integral in Eq.~(\ref{eq:KramersKronig}),
$\Lambda'=2\sqrt{\Lambda^2+m^2}$, 
must be chosen to be the same as that 
used in the evaluation of the static quantities \cite{Hatsuda:1994pi}, 
which ensures that $D_\sigma^R(\bm{q},q_0)$ at small $\bm{q}$ and $q_0$ 
diverges at the critical point of second order phase 
transition determined in the MFA \cite{KKN06}.

The imaginary parts of $Q^R_{\sigma}(\bm{q},q_0)$ and
$Q^R_{\pi}(\bm{q},q_0)$ are proportional to the difference 
between the decay and creation rates of each mode.
It is easily shown that ${\rm Im}Q^R_{\sigma(\pi)}(\bm{q},q_0)$
take nonzero values for $ |q_0| > \sqrt{ q^2 + 4m^2 }$
and $ |q_0| < q $, with $ q = |\bm{q}| $.
The decay process for $ q_0 > \sqrt{ q^2 + 4m^2 }$ in each
channel corresponds to that into a quark and an antiquark,
while the one in the spacelike region 
represents the Landau damping.

Collective modes in the $\pi$ channel are characterized
by the poles of the propagator $D^R_{\pi}(\bm{q},q_0)$.
When a pole is on the real axis, its location, 
$q_0 = \omega_\pi(q)$, i.e., the dispersion relation of 
the bound pionic modes, is determined by solving 
\begin{align}
  {\rm Re} [ D_\pi^R ( \bm{q},\omega_\pi(q) ) ]^{-1}
  = - \frac1{2G_S} - {\rm Re} Q^R_\pi(\bm{q},\omega_\pi(q) )
  = 0 ,
  \label{eq:D=0}
\end{align}
with the residue $Z_\pi(q)$ of the pole 
\begin{align}
  \frac1{Z_\pi(q)} &= -\frac1\pi
  \frac{\partial [D_{\pi}^R(\bm{q},\omega_\pi(q))]^{-1}}
  {\partial q_0}\Big{\vert}_{q_0=\omega_{\pi}(q)} \nonumber \\
  &= -\frac1\pi
  \frac{\partial Q_{\pi}^R(\bm{q},\omega_\pi(q))}
  {\partial q_0}\Big{\vert}_{q_0=\omega_{\pi}(q)}.
\end{align}
The pole on the real axis can exist in the range 
$ q < |\omega_\pi(q)| < \sqrt{ q^2 + 4m^2 }$ in which 
${\rm Im}Q_{\pi}^R ( \bm{q},q_0 )$ vanishes.
While a solution of Eq.~(\ref{eq:D=0}) no longer 
corresponds to a bound pole when $\omega_\pi(q)$ 
is outside this range, it is known that $\omega_\pi(q)$ 
approximately represents the real part of the 
corresponding pole on the lower-half complex-energy plane.

In the vacuum, $D^R_{\pi}(\bm{q},q_0)$ has a bound pole 
on the real axis as the pseudo-Nambu-Goldstone pion.
As $T$ is raised, the pionic modes eventually become unstable
against the decay into a quark and an antiquark,
as the constituent quark mass $m$ becomes smaller 
while the rest mass of pions, $\omega_\pi(0)$, 
becomes larger as shown in Fig.~\ref{fig:mass}
\cite{Hatsuda:1994pi}.
We denote the temperature at which the rest pionic modes 
become unstable by $T_{\rm ZB}$
and call it the pion zero-binding temperature.
Since the rest pionic modes are unstable for 
$\omega_\pi(0)>2m$, $T_{\rm ZB}$ is determined by 
\begin{align}
  [ D_\pi^R ( \bm{0}, 2m ) ]^{-1}_{T=T_{\rm ZB}} = 0.
\end{align}
In our model, the dissociation takes place at 
$T_{\rm ZB}=216$ MeV which is depicted in 
Fig.~\ref{fig:mass} by the vertical line.
Note that the value of $T_{\rm ZB}$ is higher than 
$T_{\rm PC}$ irrespective of the choices of the 
definition discussed before.
In Fig.~\ref{fig:mass}, the solution of Eq.~(\ref{eq:D=0})
for the $\sigma$ channel with $q=0$, $\omega_\sigma(0)$, is also 
shown. 
As in the figure, the solution is always in the continuum,
i.e., $\omega_\sigma(0)>2m$, which means that the stable $\sigma$ 
mode does not exist in our model \cite{Hatsuda:1994pi}.

\begin{figure}[t]
\includegraphics[width=0.49\textwidth]{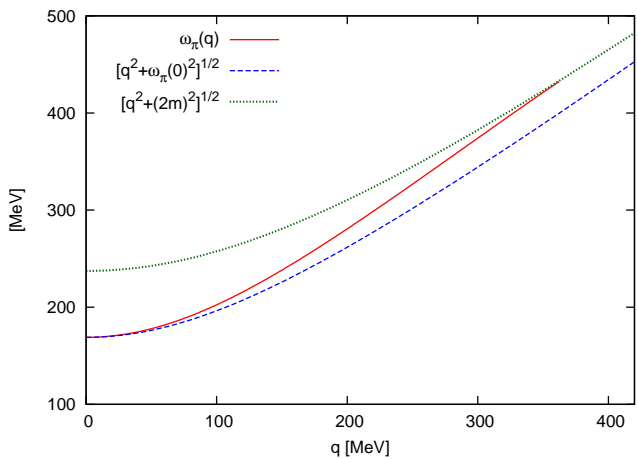}
\caption{
The dispersion relations of the pionic modes (solid line) and 
the relativistic dispersion relation for free particles
$\sqrt{q^2+[\omega_\pi(0)]^2}$ (dashed line) at $T=206$ MeV. 
The dotted line denotes the continuum threshold 
$\sqrt{q^2+4m^2}$.}
\label{fig:disp}
\end{figure}

In the vacuum, the dispersion relation of the pions 
should obey the relativistic one 
\begin{align}
\omega_{\rm rel}(q) = \sqrt{q^2 + [\omega_\pi(0)]^2},
\label{eq:KG}
\end{align}
because of the Lorentz symmetry.
In the medium at nonzero temperature and/or baryonic density, 
however, $\omega_\pi(q)$ can deviate 
from this form since the medium effect violates 
the Lorentz symmetry \cite{Shuryak:1990,Pisarski:1996}.
In Fig.~\ref{fig:disp} we show the dispersion relation of 
the bound pionic modes $\omega_\pi(q)$ at $T=206$ MeV,
which is slightly below $T_{\rm ZB}$.
One finds that $\omega_\pi(q)$ clearly deviates from 
the Lorentz-invariant form shown by the dashed line in the figure.
It is also notable that $\omega_\pi(q)$ enters the continuum
and the pionic modes become unstable at $q\simeq360$ MeV.
This result indicates that a pionic mode moving with a 
large velocity relative to the medium can become unstable 
even when the rest pion can exist as a bound state.
As we will see later, the deviation of $\omega_\pi(q)$
from the relativistic form plays a crucial role for the 
emergence of the unexpected behaviors of the quark 
spectrum for $T\lesssim T_{\rm ZB}$.

The result in Fig.~\ref{fig:disp} shows that the pion 
dispersion relation near $T_{\rm ZB}$ is steeper than 
the relativistic one, Eq.~(\ref{eq:KG}).
We note that this result in our model is to some extent
affected by the explicit breaking of Lorentz symmetry 
due to the three-dimensional cutoff besides the genuine medium effect.
In fact, the dispersion relation $\omega_\pi(q)$ in our 
model slightly deviates from Eq.~(\ref{eq:KG}) toward
steeper direction even in the vacuum.
Nonetheless, as shown in Ref.~\cite{Hatsuda:1994pi},
there are some advantages to adopt
this cutoff, and 
we can show that the resultant van Hove singularity near
the pseudocritical point appears irrespective of the
cutoff scheme as mentioned in Sec.IV B.

The qualitative structure of the pion dispersion relation 
in the medium has been discussed in various contexts 
\cite{Shuryak:1990,Pisarski:1996}.
Among them, it is shown in Ref.~\cite{Pisarski:1996} 
that the pion dispersion relation 
at sufficiently low temperature becomes 
shallower than in the vacuum on the basis of the chiral symmetry 
and Nambu-Goldstone nature of the pions.
On the other hand, it seems that
there is no conclusive argument on
the behavior of the dispersion relation at $T$
above the pseudocritical temperature.
Because the structure of the pion dispersion relation
plays a crucial role on the 
quark spectrum, we will come back to this point 
later in Sec.~\ref{sec:num}.

Before closing this section, we introduce the 
spectral function of the sigma (pionic) mode
\begin{align}
  \rho_{\sigma(\pi)} ( \bm{q},q_0 )
  = -\frac1\pi {\rm Im} D_{\sigma(\pi)}^R ( \bm{q},q_0 ) .
 \label{eq:spc-sp}
\end{align}
When $D_\pi^R ( \bm{q},q_0 )$ has a bound pole, 
$\rho_\pi ( \bm{q},q_0 )$ is decomposed as
\begin{align}
  \rho_\pi ( \bm{q},q_0 )
  = \rho_\pi^{\rm cont} ( \bm{q},q_0 )
  + \rho_\pi^{\rm pole} ( \bm{q},q_0 ) ,
  \label{eq:rho_pi}
\end{align}
where $\rho_\pi^{\rm cont}( \bm{q},q_0 )$ is the continuum part 
taking nonzero values for $ |q_0| > \sqrt{ q^2 + 4m^2 }$
and $ |q_0| < q $, and 
\begin{align}
  \rho_\pi^{\rm pole} ( \bm{q},q_0 ) 
  = Z_\pi(q) \left[ \delta( q_0 - \omega_\pi(q) )
  - \delta( q_0 + \omega_\pi(q) ) \right ].
  \label{eq:rho_pole}
\end{align}

\section{Quark spectral function}
\label{sec:quark}

The collective modes
composed of quarks and antiquarks have a natural coupling with quarks, which
in turn leads to a modification of the spectral properties of quarks, 
in particular, near the pseudocritical temperature.
To show how this modification is significant,
let us calculate the quark propagator coupled with the sigma and pionic modes in the 
random phase approximation \cite{KKN06}. 
The quark self-energy in the imaginary time formalism
in this approximation is given by
\begin{align}
  \tilde{\Sigma}(\bm{p}=0, \omega_n)
  &\equiv \tilde{\Sigma}(\omega_n)
    = -T \sum_m \int \frac{d^3q}{(2\pi)^3} 
    \nonumber \\ &
  \times \{
      \mathcal{D}_\sigma(\bm{q}, \omega_n-\omega_m) \mathcal{G}_0(\bm{q}, \omega_m)
      \nonumber \\ &
    +3\mathcal{D}_\pi(\bm{q}, \omega_n-\omega_m)
      i\gamma_5 \mathcal{G}_0(\bm{q}, \omega_m)i\gamma_5 \},
  \label{eq:sgtld}
\end{align}
with the Matsubara propagators of the sigma and pionic modes
$\mathcal{D}_{\sigma(\pi)}(\bm{q}, \nu_n)$.
The quark propagator in this approximation is diagrammatically
represented in Fig.~\ref{fig:diagram}.
The factor 3 in the second term 
in Eq.~\eqref{eq:sgtld} comes from the isospin degeneracy of pions.
Since we are interested in excitation modes at low energy 
and low momentum, 
we limit our attention to the quark spectrum at zero momentum.
The summation of the Matsubara frequency
 can be carried out analytically with
an equivalent contour integral on the complex energy plane
\cite{Kitazawa:2005vr}.
Then, after the analytic continuation
$i\omega_n\to p_0+i\eta$, 
we obtain the retarded quark self-energy,
\begin{align}
  \Sigma^R(p_0)
  &= \Sigma^R_{\sigma}(p_0) + \Sigma^R_{\pi}(p_0) ,
  \label{eq:Sigma^R}
\\
\Sigma^R_\sigma( p_0)
   &= \frac{1}{2}\sum_{s=\pm} \int\frac{d^3q d\omega}{(2\pi)^4}
      \frac{ \pi \rho_\sigma(\bm{q}, \omega)}{\omega-p_0+sE_q-i\eta}
\nonumber \\ &
    \times  \left(\gamma^0+s\frac{m}{E_q}\right)
      \left[ \coth\left(\frac{\omega}{2T}\right) + \tanh\frac{sE_q}{2T} \right],
      \label{eq:Sigma_s} \\
\Sigma^R_\pi(p_0)
  &= \frac{1}{2} \sum_{s=\pm} \int\frac{d^3q d\omega}{(2\pi)^4}
     \frac{3 \pi \rho_\pi(\bm{q}, \omega)}{\omega-p_0+sE_q-i\eta}
\nonumber \\ &
      \times \left(\gamma^0-s\frac{m}{E_q}\right)
      \left[ \coth\left(\frac{\omega}{2T}\right) + \tanh\frac{sE_q}{2T} \right],
  \label{eq:Sigma_p}
\end{align}
with $E_q=\sqrt{\bm{q}^2+m^2}$.
To avoid the ultraviolet divergence in 
Eqs.~(\ref{eq:Sigma_s}) and (\ref{eq:Sigma_p}), 
we first determine the imaginary part that is free from 
the divergence, and then evaluate the real 
part with the Kramers-Kronig relation 
\begin{align}
  {\rm Re}\Sigma^R_{\sigma(\pi)}(p_0) = 
  -\frac{1}{\pi}{\rm P}\int_{-\Lambda}^{\Lambda} dp_0' 
  \frac{{\rm Im}\Sigma^R_{\sigma(\pi)}(p_0')}{p_0-p_0'} ,
\label{eq:KramersKronig2}
\end{align}
where the energy integral is regularized by 
the cutoff $\Lambda$ \cite{KKN06}.

The retarded quark propagator for zero momentum,
\begin{align}
  G^R( p_0 ) 
  = \frac1{ (p_0 + i\eta ) \gamma^0 - m - \Sigma^R (p_0) } ,
  \label{eq:G^R}
\end{align}
is decomposed in terms of the projection operators 
$\Lambda_\pm=(1\pm\gamma_0)/2$ as 
\begin{align}
  G^R( p_0 ) 
  = G_+ (p_0) \Lambda_+ \gamma^0 + G_- (p_0) \Lambda_- \gamma^0,
  \label{eq:G^RL}
\end{align}
with 
\begin{align}
  G_\pm ( p_0 ) 
  = \frac12 {\rm Tr} [ G^R \gamma^0 \Lambda_\pm ]
  = \frac1{ p_0 + i\eta \mp m - \Sigma^\pm ( p_0 ) },
  \label{eq:Gpm}
\end{align}
and $\Sigma^\pm ( p_0 ) 
= (1/2) {\rm Tr} [ \Sigma^R(p_0) \Lambda_\pm \gamma^0 ]$
\cite{KKMN}.
The quasiquark and quasi-antiquark spectral functions
are defined in accordance with Eq.~(\ref{eq:G^RL}) as 
\begin{align}
  \rho_\pm(p_0)=-\frac{1}{\pi}\textrm{Im}G_\pm ( p_0 ) .
  \label{spct}
\end{align}
For vanishing quark chemical potential, 
the charge conjugation symmetry of the medium ensures 
the symmetry relation $\rho_-(p_0)=\rho_+(-p_0)$.
In the analysis of the quark spectrum in the next 
section, we thus concentrate on $\rho_+(p_0)$.

When the pionic modes have a bound pole,
$\Sigma^\pm(p_0)$ are decomposed as 
\begin{align}
\Sigma^\pm(p_0)=\Sigma^\pm_\sigma(p_0)
+\Sigma^\pm_{\pi\textrm{-pole}}(p_0)
+\Sigma^\pm_{\pi\textrm{-cont}}(p_0),
\label{eq:Sigma_decomp}
\end{align}
where $\Sigma^\pm_\sigma(p_0)$ represents the contribution of 
Eq.~(\ref{eq:Sigma_s}). The contribution of the pionic modes, 
Eq.~(\ref{eq:Sigma_p}), is decomposed into those of 
$\rho_\pi^{\rm cont}(\bm{q},p_0)$ and 
$\rho_\pi^{\rm pole}(\bm{q},p_0)$ in Eq.~(\ref{eq:rho_pi}).
Using Eq.~(\ref{eq:rho_pole}), one obtains 
\begin{align}
  & {\rm Im}\Sigma^\pm_{\pi\textrm{-pole}}(p_0) 
  = \frac3{4\pi^2} \sum_{r,s=\pm} \sum_{q=q_{sr}} 
  \big[
Z_\pi(\omega(q))  \left(1\mp r \frac{m}{E_q} \right)
  \nonumber \\ &\qquad
\times q^2\left|\frac{d{\cal E}_s(q)}{dq}\right|^{-1}
  \left[ 1+n(r\omega(q))-f(s E_q) \right] \big] ,
  \label{eq:imsigpole}
\end{align}
where $q_{sr}$ with $r,s=\pm$ are solutions of 
\begin{align}
  p_0 = r {\cal E}_s (q_{sr}) = r ( s E_q + \omega_\pi(q) )_{q=q_{sr}} ,
  \label{eq:q_st}
\end{align}
with ${\cal E}_\pm(q) = \pm E_q + \omega_\pi(q)$, and 
the sum in Eq.~(\ref{eq:imsigpole}) is taken for all 
the solutions of Eq.~(\ref{eq:q_st}) for each $s$ and $r$.
The functions $n(x)$ and $f(x)$ are the Bose-Einstein 
and the Fermi-Dirac distribution functions,
$ n(x) = [ \exp(x/T) - 1 ]^{-1} $ and 
$ f(x) = [ \exp( x/T) + 1 ]^{-1}$, respectively.
The product of the first two factors in the second line 
in Eq.~\eqref{eq:imsigpole}
is proportional to the difference between  the quark and pion-mode density of states,
which is called the joint density of states.
We note that 
 $d{\cal E}_{-}(q)/dq = -dE_q/dq+d\omega_{\pi}(q)/dq$
has the meaning of the relative group velocity of the quark and the
pion mode \cite{Shuryak:1990,Pisarski:1996}.

When $\omega_\pi(q)$ is of the relativistic (hyperbolic) form 
as given by Eq.~(\ref{eq:KG}), ${\cal E}_\pm(q)$ are monotonic 
functions of $q$. Equation~(\ref{eq:q_st}) thus 
can have at most one solution for a given $p_0$.
In the hot and dense medium, however, the deviation of $\omega_\pi(q)$ 
from the hyperbolic form can provide multiple 
solutions of these equations for a given $p_0$.
The nonhyperbolic form of $\omega_\pi(q)$ can 
also lead to zeros of the relative group velocity $d{\cal E}_-(q)/dq$.
For energies $p_0$ where $d{\cal E}_s(q_{sr})/dq$ vanish, 
${\rm Im}\Sigma^\pm_{\pi\textrm{-pole}}(p_0)$ diverges 
owing to the divergence of the joint density of 
states $q^2 [d{\cal E}_-(q)/dq]^{-1}$ with $q\ne0$.
Such singularities are known as the van Hove singularity
\cite{vanHove,CM,Braaten:1990wp}.
As we will see in the next section, the van Hove 
singularities in ${\rm Im}\Sigma^\pm_{\pi\textrm{-pole}}(p_0)$,
which manifest themselves as a consequence of the 
composite nature of bound pionic modes and medium effects, 
plays a crucial role to modify the quark spectral function 
significantly for $T \lesssim T_{\rm ZB}$.

\section{Numerical results}
\label{sec:num}

\subsection{Near pseudocritical temperature}
\label{sec:num1}

\begin{figure}[t]
\includegraphics[width=0.49\textwidth]{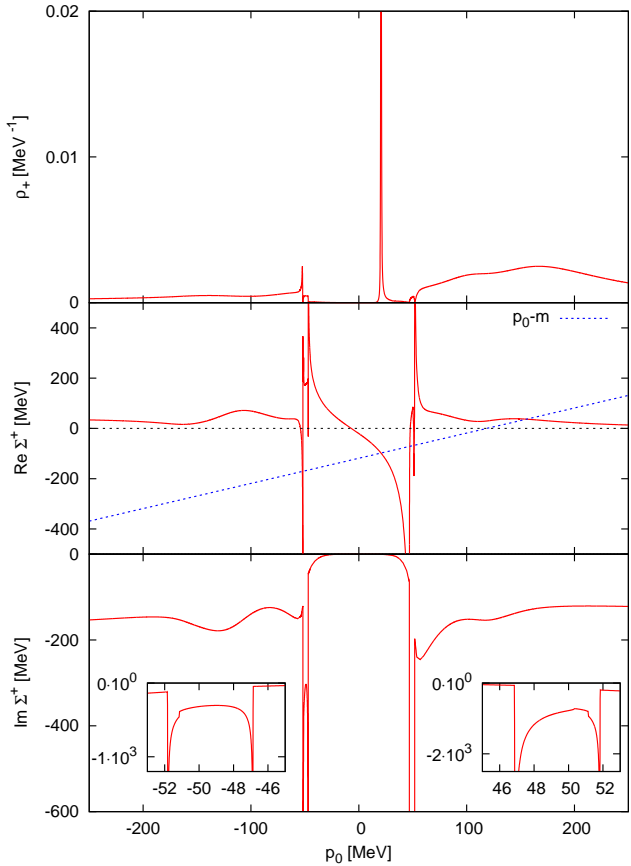}
\caption{
The upper panel shows the quark spectral function $\rho_+(p_0)$
for $T=206$ MeV.
The middle and lower panels represent the real and imaginary 
parts of the corresponding quark self-energy $\Sigma^+(p_0)$.
The dashed line in the middle panel denotes $p_0-m$.}
\label{fig:spc10}
\end{figure}

Now we present the numerical results for the quark spectrum.
In this subsection, we first investigate the effects of 
the bound pionic modes below $T_{\rm ZB}$ on the quark spectrum.
To see this effect, we fix the
temperature to $T=206$ MeV throughout this subsection;
this value is chosen as a typical
temperature below $T_{\rm ZB}$ but not less than $T_{\rm PC}$.
As discussed in Sec.~\ref{sec:fluctuation}, 
the dynamic chiral susceptibility has the 
largest peak in the spacelike region at this temperature \cite{KKNprep}.
As we will see in this subsection, however, the effect 
of the sigma mode does not have a significant contribution 
to the quark spectrum even for this temperature.

We first show $\rho_+(p_0)$ at $T=206$ MeV in the upper 
panel of Fig.~\ref{fig:spc10}.
The figure shows that the quark spectrum 
is significantly modified from the one in the MFA, 
$\rho_+(p_0) = \delta(p_0-m)$, with $m\simeq120$ MeV 
being the constituent quark mass in the MFA at this temperature.
The quasiquark spectrum has a sharp peak at 
$p_0\simeq25$ MeV, which is considerably smaller than $m$.
The spectral weight of this peak may be defined by
\begin{align}
Z = \int_{\Delta} dp_0 \rho_+(p_0) ,
\end{align}
where $\Delta$ is a range of $p_0$ that well covers the peak structure.
The numerical calculation gives $Z\simeq0.16$, which is 
small but not negligible.
The spectrum $\rho_+(p_0)$ also has a 
broad peak structure around $p_0\simeq160$ MeV.
While there exists another peak at $p_0\simeq-50$ MeV,
the spectral weight of this peak is negligibly small.

To understand the origin of these structures in 
$\rho_+(p_0)$, the real and imaginary parts of 
$\Sigma^+(p_0)$ are shown in the middle and 
lower panels of Fig.~\ref{fig:spc10}, respectively.
The real part of $\Sigma^+(p_0)$ 
determines the quasipoles of the quark, 
where the real part of the inverse propagator vanishes as
\begin{align}
{\rm Re}[G_+(p_0)]^{-1}
= p_0 - m - {\rm Re} \Sigma^+(p_0) = 0.
\label{eq:QuasiDisp}
\end{align}
The quasipole gives approximate position of 
a peak in $\rho_+(p_0)$ when Im$\Sigma^+(p_0)$ 
is small there \cite{KKN06}.
The solutions of Eq.~(\ref{eq:QuasiDisp}) are graphically 
determined by crossing points of ${\rm Re}\Sigma^+(p_0)$
and a line $p_0-m$ which is drawn by the dashed line
in the middle panel in Fig.~\ref{fig:spc10}.
One finds that there exists a quasipole at 
$p_0\simeq25$ MeV corresponding to the sharp peak in $\rho_+(p_0)$.
There also exists a quasipole around $p_0\simeq150$ MeV, 
but a clear peak corresponding to this quasipole 
does not appear in $\rho_+(p_0)$ 
because of the large ${\rm Im}\Sigma^+(p_0)$ around this energy.
Although there are some more solutions of Eq.~(\ref{eq:QuasiDisp})
around $p_0 = \pm 50$ MeV owing to the singular behaviors of 
${\rm Re}\Sigma^+(p_0)$, clear peaks corresponding to
these quasipoles are not formed in $\rho_+(p_0)$.

\begin{figure}[t]
\includegraphics[width=0.49\textwidth]{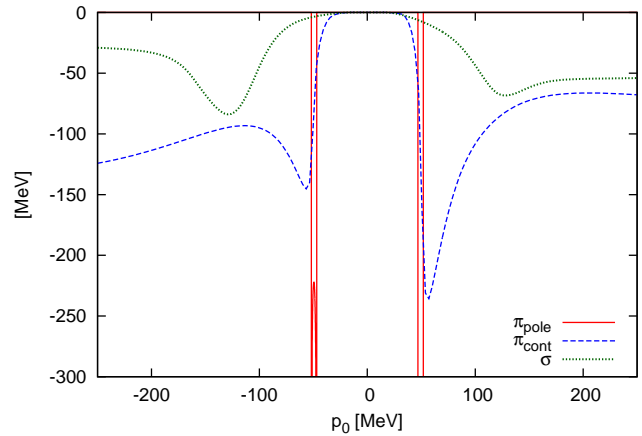}
\caption{
Decomposition of ${\rm Im}\Sigma^+(p_0)$ into three parts,
${\rm Im}\Sigma^+_\sigma(p_0)$ (solid line),
${\rm Im}\Sigma^+_{\pi\textrm{-pole}}(p_0)$ 
(dotted line), and
${\rm Im}\Sigma^+_{\pi\textrm{-cont}}(p_0)$ 
(dashed line) at $T=206$ MeV.
}
\label{fig:imsig_decomp}
\end{figure}

\begin{figure}[t]
\includegraphics[width=0.49\textwidth]{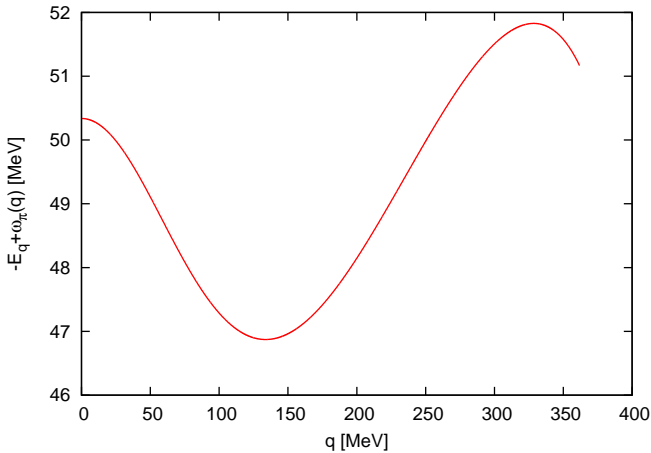}
\caption{
Momentum dependence of ${\cal E}_-(q) = -E_q + \omega_\pi(q)$
at $T=206$ MeV.
}
\label{fig:Eq-w}
\end{figure}

In the lower panel of Fig.~\ref{fig:spc10}, 
one finds that ${\rm Im}\Sigma^+(p_0)$
is divergent at four energies, 
$p_0 = \pm p_0^{(1)}$ and $\pm p_0^{(2)}$
with $p_0^{(1)}\simeq47$MeV and $p_0^{(2)}\simeq52$MeV.
As shown in the small windows in the panel, 
$|{\rm Im}\Sigma^+(p_0)|$ is large 
in the range $p_0^{(1)} < |p_0| < p_0^{(2)}$.
Through the Kramers-Kronig relation 
Eq.~(\ref{eq:KramersKronig2}),
this structure in ${\rm Im}\Sigma^+(p_0)$ in turn 
brings about the singularities in ${\rm Re}\Sigma^+(p_0)$
at $p_0=\pm p_0^{(1)}$ and $\pm p_0^{(2)}$.
These divergences  are thus responsible for the emergence 
of the quasipoles discussed above, and hence the sharp 
peak at $p_0\simeq25$ MeV in $\rho_+(p_0)$.

To clarify the origin of the divergences in 
${\rm Im}\Sigma^+(p_0)$, we show each part of 
${\rm Im}\Sigma^+(p_0)$ in the decomposition of 
Eq.~(\ref{eq:Sigma_decomp}) in Fig.~\ref{fig:imsig_decomp}.
The figure shows that the divergences come from 
${\rm Im}\Sigma^+_{\pi\textrm{-pole}}(p_0)$, i.e.,
scattering of quarks with the bound pionic modes.
As is seen from Eq.~(\ref{eq:imsigpole}),
this term takes nonzero values for $p_0$ 
satisfying Eq.~(\ref{eq:q_st}).
For $s=+1$, ${\cal E}_+(q)$ is a monotonically increasing
function of $q$ with the minimum 
${\cal E}_+(0) = m+\omega_\pi(0) \simeq 270$ MeV at $q=0$.
Equation~(\ref{eq:q_st}) thus has one solution for
$r p_0 > m+\omega_\pi(0)$ with $r=\pm1$, which, 
however, is outside the range of $p_0$ shown in 
Fig.~\ref{fig:imsig_decomp}.
With $s=-1$, on the other hand, 
${\cal E}_-(q)=-E_q + \omega_\pi(q)$ is not monotonic as shown 
in Fig.~\ref{fig:Eq-w}, and the range of ${\cal E}_-(q)$ 
is limited to $p_0^{(1)} < p_0 < p_0^{(2)} $.
Therefore, ${\rm Im}\Sigma^+_{\pi\textrm{-pole}}(p_0)$ 
takes nonzero values for $p_0^{(1)} < |p_0| < p_0^{(2)} $.
Note that the line of ${\cal E}_-(q)$ in Fig.~\ref{fig:Eq-w}
terminates around $q=360$ MeV, because $\omega_\pi(q)$ 
enters the continuum and the bound pole disappears at 
this momentum as shown in Fig.~\ref{fig:disp}.
At the extrema of ${\cal E}_-(q)$, the relative group
velocity $d{\cal E}_-(q)/dq$ vanishes. 
This leads to the divergence of the joint density of states
$q^2|d{\cal E}_-(q)/dq|^{-1}$ and the singularity of 
${\rm Im}\Sigma^+_{\pi\textrm{-pole}}(p_0)$ 
at $|p_0|= p_0^{(1)}$ and $p_0^{(2)}$.
The divergences in ${\rm Im}\Sigma^+_{\pi\textrm{-pole}}(p_0)$ 
thus come from van Hove singularity owing to the 
divergence of the joint density of states.

We remark that the van Hove singularity discussed here 
does not manifest itself if $\omega_\pi(q)$ takes the 
relativistic form Eq.~(\ref{eq:KG}), since ${\cal E}_-(q)$ 
is then a monotonic function of $q$ and $d{\cal E}_-(q)/dq$ 
remains nonzero for $q\ne0$ \cite{KKMN}.
Therefore, the van Hove singularity does not appear in the 
models composed of a fermion and boson with the hyperbolic 
dispersion relation assumed in Refs.~\cite{BBS,KKN07,KKMN}.
The composite nature and medium effects that lead to a 
nonhyperbolic form of $\omega_\pi(q)$ play a crucial 
role for realizing the van Hove singularity, and the 
drastic modification of the quark spectrum
as a result of the singularity.

For the $\sigma$ contribution, we see that
there are small peaks in Im$\Sigma_\sigma^+(p_0)$ at $|p_0|\simeq 130$ MeV.
They come from the scattering of quarks with the
sigma mode in the timelike region.
Because this mode is always in the continuum, it does not 
form a singular structure in Im$\Sigma_\sigma^+(p_0)$
unlike the pionic mode.
On the other hand, as discussed above, the dynamic chiral susceptibility
that lies in the spacelike region in the sigma mode has
the maximum at $T=206$ MeV.
This contribution to the quark spectrum is, however, so small 
that it does not lead to a peak in Im$\Sigma_\sigma^+(p_0)$.
This is due to the fact that the critical point at which
the chiral susceptibility diverges is far from this temperature and (zero)
density in this model; it is located at $T\simeq 47$ MeV and 
the quark chemical potential $\mu\simeq 329$ MeV.
Therefore, at the vanishing quark chemical potential, 
the contribution of the soft mode associated with this critical point is
negligible at any temperature.
Effects of the soft mode on the quark spectrum near the critical point will be
investigated in Ref.~\cite{KKNprep}.

\subsection{Discussion} 

Here we shall closely examine the origin of the 
sharp peak in the quark spectrum in the far-soft region in terms of
the van Hove singularity.

We first compare the quark spectrum in Fig.~\ref{fig:spc10} 
with the results in Refs.~\cite{BBS,KKN07,KKMN} where
the fermion spectra are computed in Yukawa models 
composed of an elementary fermion and boson with masses $m_f$ and $m_b$.
Since the present analysis deals with the constituent 
quarks with $m\simeq120$ MeV coupled to the  bound pions 
with the rest masses $\omega_\pi(0)\simeq150$ MeV and 
a vanishing width, the resultant quark spectrum may 
well be compared to the one obtained for the elementary 
particle systems with $m_f=m$ and $m_b=\omega_\pi(0)$.
In Ref.~\cite{KKN07}, it is found that the spectrum of 
a massless fermion coupled with a massive boson has 
a three-peak structure with a sharp peak at the origin for $T\simeq m_b$.
For a massive fermion, the peak position shifts 
toward nonzero positive energy, and gradually ceases 
to exist \cite{KKMN}; the range of $m_f$ where 
the clear peak structure is realized is limited for $m_f/m_b \lesssim 0.2 $.
Now since the present mass ratio $m/\omega_\pi(0) \simeq 0.8$
is significantly larger than this upper limit, 
the quark spectral function should never have the peak structure, 
if the boson were described as an elementary particle 
with the free dispersion relation. 
It is also notable that the clear peak in Fig.~\ref{fig:spc10} 
appears at an unexpectedly low energy, $p_0\ll m$.

As already noted in the previous sections, the crucial difference of the present analysis 
from the ones in Refs.~\cite{BBS,KKN07,KKMN} is the compositeness
of the bosonic modes with a nonhyperbolic dispersion relation 
due to the medium effect.
Owing to the modified dispersion relation 
the van Hove singularity emerges in Im$\Sigma^+(p_0)$, 
which significantly modifies the quark spectrum.
On the other hand, the collective excitation 
corresponding to the quasipole at 
$p_0\simeq150$ MeV   no longer makes a sharp peak
because of the large decay rate.
Here, the composite nature of the parasigma and parapion
is again responsible for this behavior, since the large 
decay rates around $p_0=150$ MeV come from the contribution 
of continuum spectra in the $\sigma$ and $\pi$ channels 
as shown in Fig.~\ref{fig:imsig_decomp}.

While we have emphasized the effect of the van Hove singularity
on $\rho_+(p_0)$, we notice that the divergence is not 
necessarily indispensable for the drastic modification of the quark spectrum.
The important feature is the existence of sharp peaks in ${\rm Im}\Sigma^+(p_0)$, 
i.e., a concentration of the decay rate of quasiquarks to some 
narrow energy regions.
Such a sharp peak in ${\rm Im}\Sigma^+(p_0)$ 
in turn makes a sharp rise and decrease in  ${\rm Re}\Sigma^+(p_0)$ through the 
Kramers-Kronig relation Eq.~(\ref{eq:KramersKronig2}), 
and thus leads to a distorted quark spectrum.
In fact, we will see in the next subsection that 
a somewhat moderate but still strong modification of the quark spectrum is 
realized even above $T_{\rm ZB}$ where the van Hove
{\em singularity} no longer exists because of the absence of the stable pionic modes.
When we incorporate the higher-order corrections,
the bound pionic modes and 
quarks acquire nonzero decay widths and the would-be van Hove 
singularity will turn into a  smeared peak.
Even in this case, the modification of the quark 
spectrum and the emergence of a peak in a far-soft region is expected if 
a sharp peak exists in ${\rm Im}\Sigma^+(p_0)$.
The modification of the quark spectrum induced by the 
scattering with a boson having a distorted dispersion 
relation, therefore, is expected to take place irrespective
of the details of the model and approximation used 
in the present analysis.

As mentioned in Sec.~\ref{sec:fluctuation}, 
the detailed form of the pion dispersion relation $\omega_\pi(q)$ 
in our model is affected by the cutoff scheme,
and so is the detailed properties such as the 
position and strength of the van Hove singularity in the quark self-energy,
although the drastic change of the quark spectrum itself 
takes place in a generic way
once the dispersion relations of the parapion and quarks 
take nonhyperbolic forms in the medium.
In fact, we have checked that the van Hove singularity
in the quark self-energy emerges at some 
temperature even if we employ different regularization
schemes in our model; while our model predicts a 
steep dispersion relation as shown in Fig.~\ref{fig:disp},
the singularity appears even with 
a shallow dispersion relation of the pionic mode.

For determining the position and the strength 
of  the van Hove singularity quantitatively, 
a precise determination of the spectral properties of the pionic mode,
including its dispersion relation and width, 
near $T_{\rm ZB}$ is necessary.
For this purpose, simulations on the lattice 
should hopefully  be helpful.

\subsection{High temperatures}

\begin{figure*}[t]
\includegraphics[width=0.99\textwidth]{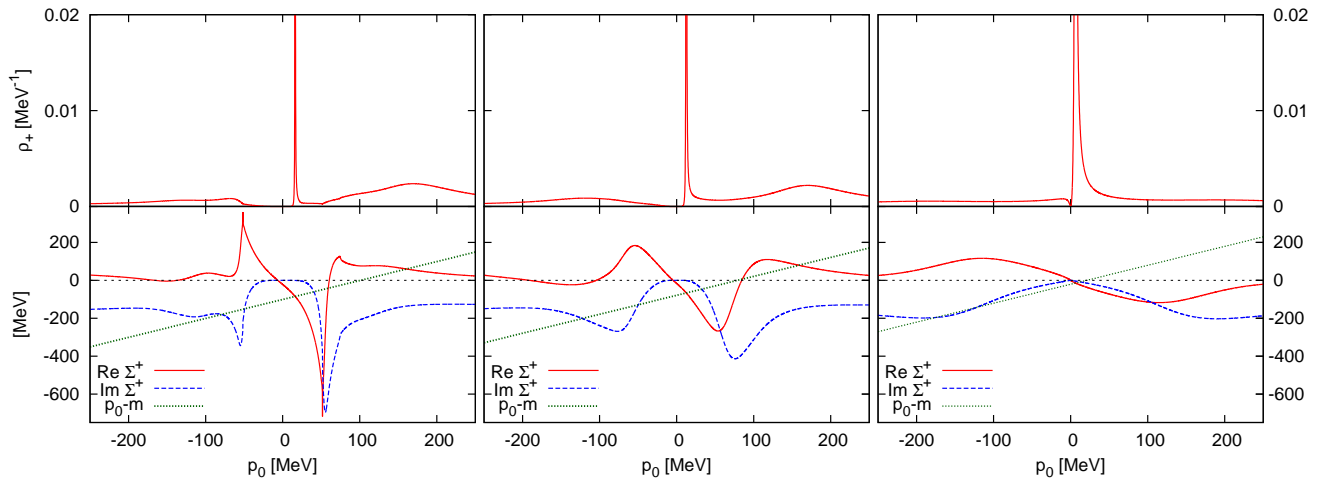}
\caption{
Quark spectrum $\rho_+(p_0)$ and corresponding 
self-energy at $T= 0.98T_{\rm ZB}$ (left), $1.02T_{\rm ZB}$
(middle) and $1.5T_{\rm ZB}$ (right).
}
\label{fig:spc9802}
\end{figure*}

Next, let us see the quark spectrum near and significantly 
above the pion zero-binding temperature $T_{\rm ZB}$.
In Fig.~\ref{fig:spc9802},
we show the quark spectrum $\rho_+(p_0)$ and 
the corresponding self-energy $\Sigma^+(p_0)$ for 
$T=0.98T_{\rm ZB}$, $1.02T_{\rm ZB}$ and $1.5T_{\rm ZB}$.
It is found from the left panel of Fig.~\ref{fig:spc9802} that
the van Hove singularity is not seen in the quark self-energy,
although the stable pionic modes still exist at $T=0.98T_{\rm ZB}$.
This is because the momentum range where the stable pionic
modes exist becomes narrow as $T$ increases and the relative
group velocity does not have a chance to vanish in the range.
We, however, see that there exist sharp but finite peaks 
in Im$\Sigma^+(p_0)$ at $|p_0|\simeq 55$ MeV.
These peaks are understood as the remnant of the van Hove 
singularity in ${\rm Im}\Sigma^+(p_0)$ in Fig.~\ref{fig:spc10}.
As a result of these peaks in ${\rm Im}\Sigma^+(p_0)$,
three quasipoles manifest themselves with the same 
mechanism discussed in the previous subsection, and 
a sharp peak is formed at low energy $p_0\simeq16$ MeV.
The position of this peak with the strength $Z\simeq0.17$ 
is much lower than the constituent quark mass 
$m\simeq100$ MeV for this temperature.

For $T>T_{\rm ZB}$, the stable pionic modes no longer exist, 
and hence the quark self-energy is smooth as a function of $p_0$.
At $T=1.02T_{\rm ZB}$, which is slightly above $T_{\rm ZB}$, 
there exist broad peaks in ${\rm Im}\Sigma^+(p_0)$ 
around $p_0\simeq\pm76$ MeV.
These peaks come from the coupling of quarks with the pionic modes;
whereas the pionic modes are no longer stable, there still 
exists a well developed collective mode 
slightly above $T_{\rm ZB}$.
As a consequence of these peaks in ${\rm Im}\Sigma^+(p_0)$,
a sharp peak is formed in $\rho_+(p_0)$ at $p_0\simeq12$ MeV
with $Z\simeq 0.18$.
The position of this peak is still considerably lower
than the constituent quark mass $m\simeq79$ MeV.
The strong modification of the quark spectrum thus 
sustains even slightly above $T_{\rm ZB}$.

As $T$ is raised further,
the bump structure in ${\rm Im}\Sigma^+(p_0)$ disappears
since the well-developed collective modes in the $\sigma$ 
and $\pi$ channels cease to exist.
This behavior is seen in the right panel of 
Fig.~\ref{fig:spc9802}, which presents the quark 
spectrum for $T=1.5T_{\rm ZB}$.
As a result, the quark spectrum approaches 
the mean field one as $T$ increases.
For $T=1.5T_{\rm ZB}$, the position of the sharp peak of 
$\rho_+(p_0)$ is close to the the constituent quark mass 
$m\simeq21$ MeV.

\section{Summary}

In the present study, we have investigated the quark 
spectrum near the pseudocritical temperature
$T_{\rm PC}$ of chiral phase transition and the pion 
zero-binding temperature $T_{\rm ZB}$ at vanishing quark chemical potential 
focusing on the effect of fluctuation modes
in the $\sigma$ and $\pi$ channels in 
the two-flavor NJL model with nonzero current quark mass $m_0$.
Compared with the previous study in the chiral limit
\cite{KKN06}, nonzero $m_0$ gives rise to the 
nonzero constituent quark mass $m$ even above $T_{\rm PC}$
because of the crossover nature of the phase transition.
In Ref.~\cite{KKMN}, 
it was shown in a Yukawa model where the boson has a dispersion 
relation valid in the free space that the nonzero fermion 
mass tends to suppress the appearance of the multipeak 
structures in the quark spectrum $\rho_+(p_0)$.
Our microscopic model calculation has shown that
$\rho_+(p_0)$ near $T_{\rm PC}$ 
is significantly modified by the scattering with stable
pionic modes that have a nonhyperbolic dispersion relation,
as was argued in various models in different contexts 
\cite{Shuryak:1990,Pisarski:1996}.
We have clarified that these modifications are 
caused by the van Hove singularity owing to the vanishing of 
the relative group velocity between quarks and bound pionic modes.
The composite nature of the pionic modes that gives rise
to the nonhyperbolic dispersion relation plays 
a crucial role for the modification of the quark spectrum.
We have found that the quark spectrum has a sharp peak 
at an energy considerably lower than the constituent 
quark mass near $T_{\rm PC}$ as a consequence of the 
van Hove singularity.

Because our results show that the quark spectrum near 
$T_{\rm PC}$ is strongly modified by the scattering with 
pionic modes, it is interesting to pursue the effects of 
this modification on other observables near $T_{\rm PC}$.
For example, the existence of light quasiquark excitation
would affect the $T$ dependence of thermodynamic observables near $T_{\rm PC}$.
It would also affect the experimental observables in heavy
ion collisions, such as the dilepton production rate \cite{Braaten:1990wp}.
Exploring the existence of the van Hove singularity in the 
early Universe in neutrino spectra and estimating their effects
on the formation of baryon asymmetry \cite{Miura:2013fxa} 
are also interesting subjects.

In this paper we have concentrated on the quark spectrum
at zero momentum and evaluated the quark self-energy at the one-loop order.
Since the medium near $T_{\rm PC}$ is thought to be a strongly correlated system,
it is more desirable to adopt a more sophisticated approximation 
 taking into account the self-consistency between the fluctuation modes and the quasiquarks, 
as was done for other problems in Ref.~\cite{Muller:2010am},
in which the investigation is, however, not for the system close to $T_{\rm PC}$,
and the van Hove singularity is not seen.
Indeed, as shown in the present work, when the system is far
from  $T_{\rm PC}$, the peak in the quark spectrum
is close to the one in the mean field approximation and the van Hove
singularity does not occur. 
It would be quite interesting to investigate the quark spectrum near
$T_{\rm PC}$ in such an approach.
Such an investigation of the quark spectrum around the pseudocritical
temperature is,
however, beyond the scope of the present work and left for a future
project.

This work is in part supported by JSPS KAKENHI
Grants No. 25800148, 
No. 20540265, 
No. 23340067, 
No. 24340054, and 
No. 24540271. 
T.K. was partially supported
by the Yukawa International Program for Quark-Hadron Sciences.


\end{document}